\documentstyle[11pt, newpasp,twoside,epsf]{article}
\def\lsim {\lower .1ex\hbox{\rlap{\raise .6ex\hbox{\hskip .3ex
        {\ifmmode{\scriptscriptstyle <}\else
                {$\scriptscriptstyle <$}\fi}}}
        \kern -.4ex{\ifmmode{\scriptscriptstyle \sim}\else
                {$\scriptscriptstyle\sim$}\fi}}}
\def\eps@scaling{.95}
\def\epsscale#1{\gdef\eps@scaling{#1}}
 \def\plotone#1{\centering \leavevmode
\epsfxsize=\eps@scaling\textwidth \epsfbox{#1}}

\begin{document}

\title{Sweeping and shaking dwarf satellites}

\author{Lucio Mayer}
\affil {Institute for Theoretical Physics, University of Z\"urich, CH-8057 Zurich, Switzerland}

\author{James Wadsley}
\affil {Department of Physics \& Astronomy, McMaster University, Canada}

\begin{abstract}

We present the first high-resolution N-Body/SPH simulations that follow
the evolution of low surface brightness disk satellites 
in a primary halo containing both dark matter and a hot gas 
component.
Tidal shocks turn the stellar disk into a spheroid with low $v/\sigma$ and
remove most of the outer dark and baryonic mass. In addition, by weakening
the potential well of the dwarf tides enhance the effect of ram pressure, and
the gas is stripped down to radius three times smaller than the stellar component.
A very low gas/stars ratio results after several Gyr, similarly to what seen in dwarf spheroidal satellites of the Milky Way and M31.

\end{abstract}

\section{Introduction}

One of the most striking properties of the Local Group (LG) 
and other nearby groups 
is the morphological segregation shown by its galaxy members. Dwarf irregulars
(dIrrs) 
are mostly found at distances  $> 300$ kpc from the Milky Way and M31,
while dwarf spheroidals (dSphs) and dwarf ellipticals (dEs) are clustered around the latter (Mateo 1998).
Tidal stirring of disk-like dwarfs once they fall into
the halo of the primaries at high redshift provides a way to 
transform their stellar component; a rotationally supported 
disk like that of dIrrs is heated by tidal shocks
and by an induced bar-buckling instability, and simultaneously loses mass, 
turning into a pressure supported  spheroidal of lower luminosity
(Mayer et al. 2001a,b, hereafter MA01a,b).
The periodic bursts of star formation seen in some dSph satellites of the 
Milky Way can also be explained as being triggered
at pericenter passages (MA01a).
However, stripping by tides plus gas consumption from star formation
yields final gas fractions an order of magnitude too high compared
to the limits inferred for dSphs (MA01b). 
A gas-poor dwarf can be the outcome of heating by the cosmic UV
background (Bullock, Kravtsov \& Weinberg 2000), but models of the
tidal mass loss (Taffoni et al. 2003) suggest that the progenitors of Fornax, 
Sagittarius or the dEs satellites of M31 were too large ($V_c > 50$ km/s) for
photoevaporation to be effective.

Recent new observational 
constraints coming from X-ray and UV
absorption measurements suggest that the Milky Way has a hot gaseous
corona with a mean density $2 \times 10^{-5}$ atoms cm$^{-3}$ within 150 kpc
(Sembach et al. 2003).
Ram pressure in a tenuous hot halo could
remove the gas of dSphs at their present densities
(Blitz \& Robishaw 2002; Gallart et al. 2001). However, the current structure of dSphs does not
reflect the initial conditions because they have been deeply affected by tides.
Here we present results
from new simulations that follow a disk-like
dwarf orbiting in a ``live'' MW-sized halo with both a dark matter and a hot
gas component, witnessing for the first time the combined
effect of tides and ram pressure.

\section{Initial Conditions}

Both the primary and the dwarf satellite are modelled according to current
galaxy formation models in the LCDM cosmogony (Mo, Mao \& White 1998).
The primary (dark+hot gas) halo is based on the Milky Way (see Mastropietro
et al., these proceedings). Its rotation curve peaks at 220 km/s.
The gaseous halo density at $50$ kpc is either $2\times 10^{-5}$ or $8 \times 10^{-5}$ atoms/cm$^{3}$, within the observational constraints. The
dwarf has a dark (NFW) halo with $V_{vir}=50$ km/s, a
stellar and a gaseous component. The gas accounts for 30\% of the total disk
mass (this is $4 \%$ of the virial mass) and has a mean density 
$\rho_g \sim 10^{-3}$ cm$^{-3}$, as found in LG dIrrs (Grebel, Gallagher
\& Harbeck 2003).

\begin{figure}
\epsscale{0.7}
\plottwo{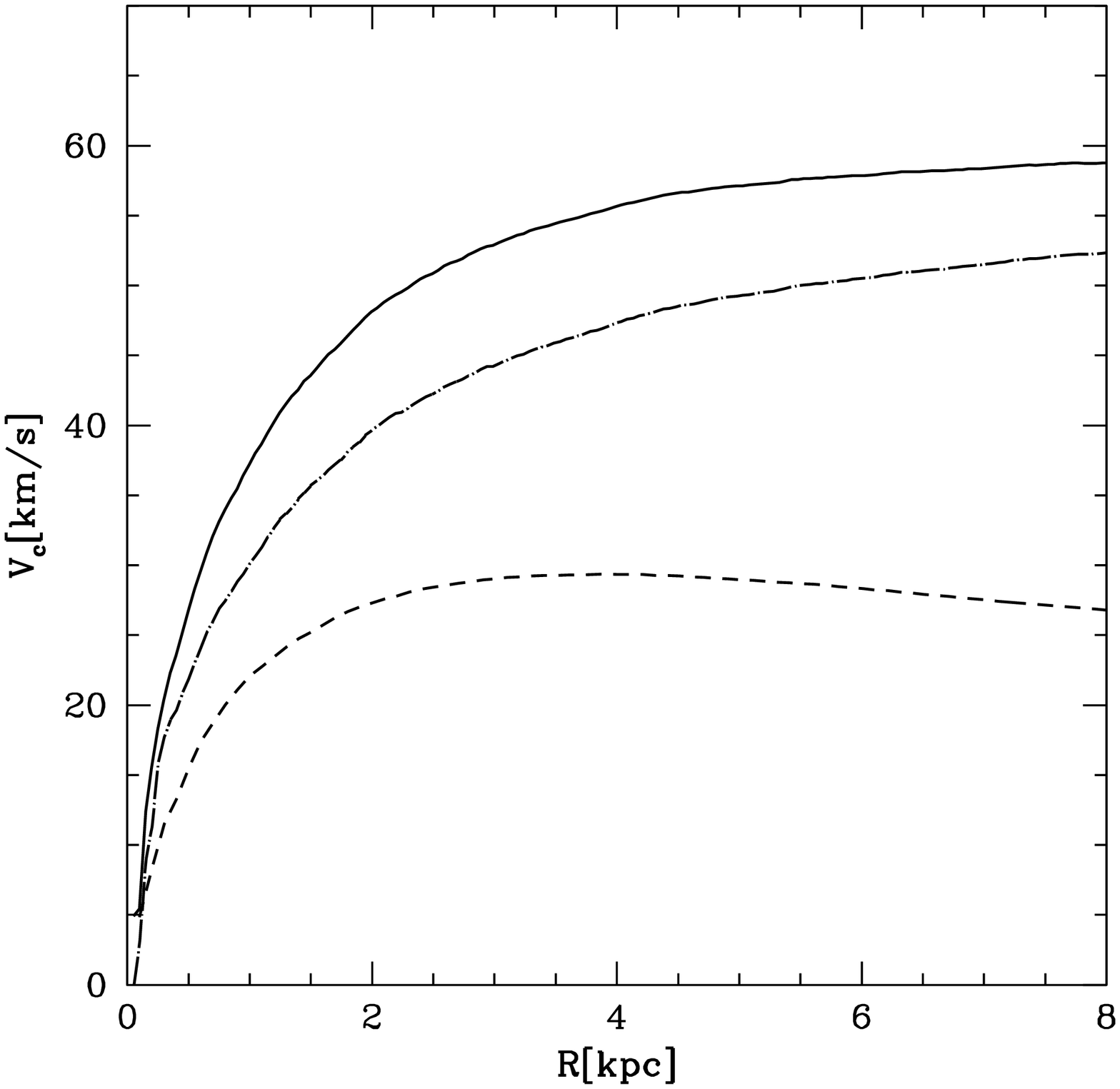}{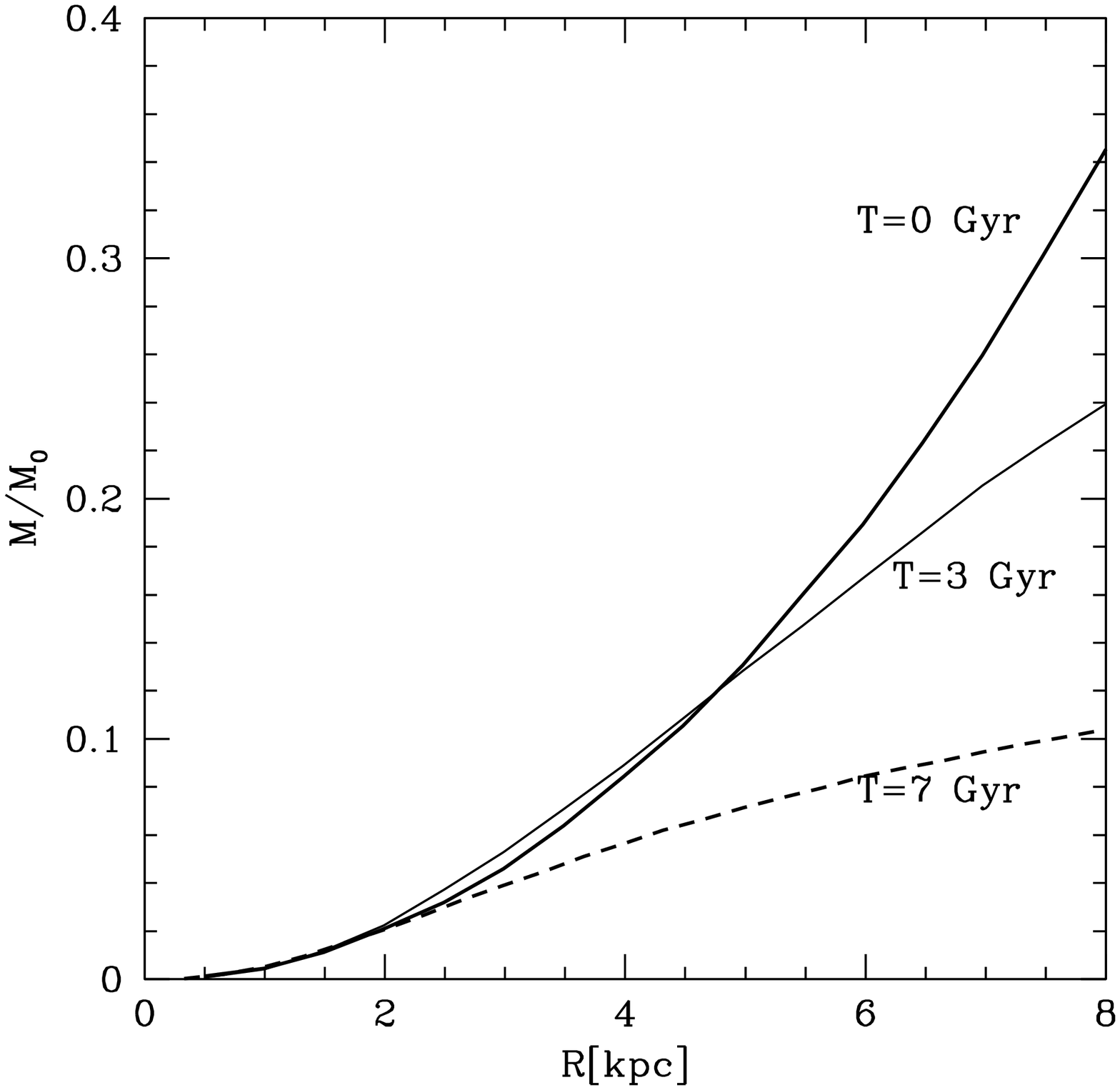}
\caption{Left: rotation curve of the initial dwarf model. The solid line shows
the global curve, the dot-dashed line the contribution of the dark matter
 and the short-dashed  that of the disk. Right: evolution of the mass profile of 
the gas (normalized to the initial total gas mass) in the run in which 
$n_H = 8 \times 10^{-5}$ atoms/cm$^{3}$ at 50 kpc.}
\label {fig1}
\end{figure}

The halo has a low concentration, $c=4$, as required
to match the rotation curves of dIrr galaxies (Swaters et al. 2003).
The spin parameter is $\lambda=0.043$, resulting in a disk scale length
of $2$ kpc, and the B band central surface brightness, assuming a
stellar mass-to-light ratio $M/L_B = 2$, is 
23 mag arcsec$^{-2}$. The orbit has a pericenter of $50$ kpc and
an apocenter of $250$ kpc, such eccentricity being typical for satellites 
in cosmological simulations. With the exception of Leo I and Leo II, the orbits of LG
dSphs are much closer to the primary galaxies compared to our choice,
hence our results on the effect of tides and ram pressure will
be on the conservative side. 
Overall the simulations use 1.85 million particles
of which most of them are in the 
primary halo to minimize two-body heating against the lighter 
particles of the dwarf (Abadi, Moore \& Bower 2000). The dwarf has 300,000
particles in the halo and 50,000 (20.000 gas and 30.000 star particles) in
the disk. The simulations are carried out with the parallel Tree+SPH code
GASOLINE (Wadsley, Stadel \& Quinn 2003) without radiative cooling.

\section{Results}

The dwarf galaxy is evolved for 7 Gyr (about 2 orbital times) in the primary
potential. Close to pericenter the gaseous disk begins to be ablated
by ram pressure (Figure 2),simultaneously being distorted and turned into a non-axisymmetric shape by the tides. A bar forms in the stellar component
which torques the gas, funnelling the latter inwards on a dynamical timescale as reported in previous simulations. The gas inflow counteracts stripping by
tides and ram pressure, momentarily increasing the gas mass within the inner 
4 kpc, while the outer disk is being completely removed (right  panel in Figure 1).

\begin{figure}
\epsscale{0.99}
\plotone{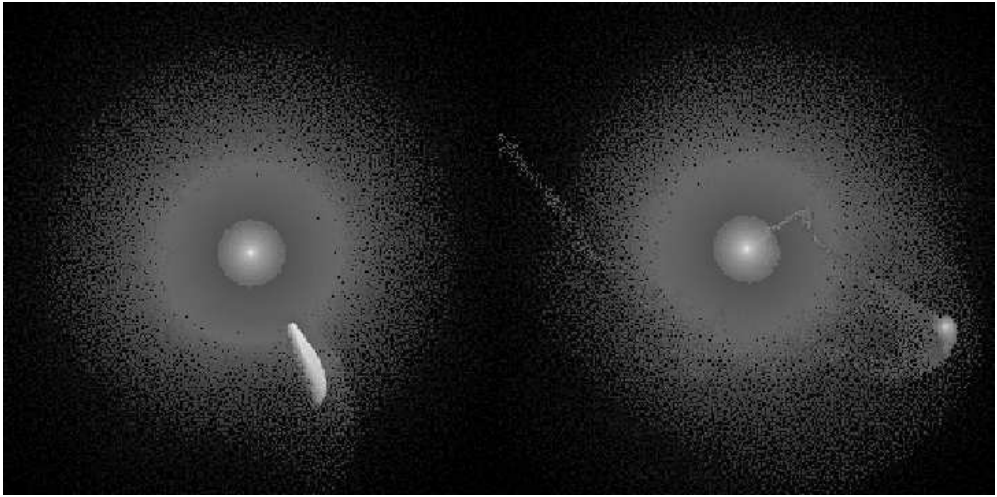}
\caption{Gas density close to first pericenter (left) and second pericenter
passage (right). Boxes are 100 kpc on a side, brighter regions correspond 
to higher densities.}
\label {fig2}
\end{figure}

After the second pericenter passage, however, ram pressure stripping has been 
efficient enough to remove the gas down to 2.5 kpc (Figure 1); such a stripping radius is almost 3 times 
smaller than that expected
from ram pressure in absence of tides (Blitz \& Robishaw 2001). At the end of the simulation
the system resembles a gas-poor pressure supported dSph;
the measured central $v/\sigma$ has dropped to less than 0.3, $\sigma \sim$ 20
km/s, between $80$ and $90$\% of the gas has been removed depending on 
the halo density, while only $30 \%$ of the stars have been stripped by tides,
and the dark matter still accounts for about $80\%$ of the mass.
The final gas mass within the gravitationally bound component of the galaxy
is 5-7\% of the mass in stars; this is roughly a factor of 4-5
lower than previous simulations which included only tides 
(MA01b). The upper limit on the HI/stars mass ratio in Local
Group  dSphs is still a bit lower, about 3\%, 
but we cannot exclude that there is as much as 10\% of the stellar mass 
in the form of a warm ionized component (Grebel, Gallagher \& Herbeck 2003).

\section{Conclusions}

These simulations show that a model with tides and ram pressure in a tenuous
hot Galactic corona is able to produce a very low gas-to-stars ratio in
a dwarf satellite after a few orbits. However, it remains to be seen if
the same model can bring down the gas fraction by another factor of ten
and match the HI/stars ratio of Draco and Ursa Minor ($\sim 0.1 \%$);
in principle this is possible because their progenitors were likely at least 
ten times  lighter and their orbit located much closer to the
Galaxy compared to the system simulated here (MA01b).
In addition radiative cooling and star formation should be included
in the simulations; however, new runs still in progress suggest that their
combined effect leaves the final gas fraction almost unchanged, as the
larger amount of gas retained by the satellite thanks to cooling compensates 
for the fraction that is turned into stars.
Instead, a clumpy ISM and diffusion through the interface between the
cold disk and the hot ambient medium induced by Raleygh-Taylor instabilities
(not captured by SPH at the current resolution) could enhance significantly
gas stripping (Quilis, Moore \& Bower 2000).


\begin{references}

\reference Abadi, M.G., Moore, B., \& Bower, R.G., 1999, MNRAS, 308, 947
\reference Blitz, L., \& Robishaw, T., 2000, ApJ, 541, 675 
\reference Bullock, J.S., Kravtsov, A.V., \& Weinberg, D.H., 2000, ApJ, 539, 517
\reference Gallart, C. Martinez-Delgado, D., Gomez-Flechoso, M.A., \& Mateo, M. 2001, AJ, 121, 2572
\reference Grebel E.K., Gallagher, J.S., \& Harbeck, D., 2003, 125, 1926   
\reference Mateo, M.L., 1998, ARA\&A, 36, 435
\reference Mayer, L., Governato, F., Colpi, M., Moore, B., Wadsley, J., Stadel,
J., \& Lake, G., 2001a, 547, L123
\reference Mayer, L., Governato, F., Colpi, M., Moore, B., Wadsley, J., Stadel,
J., \& Lake, G., 2001b, 559, 754
\reference Mo, H.J., Mao, S., \& White, S.D.M., 1998, MNRAS, 295, 319
\reference Quilis, V., Moore, B., \& Bower, R., 2000, Science, 288, 1617
\reference Sembach, K. R. et al. 2003, ApJS, 146, 165 
\reference Swaters, R.A., Madore, B.F., Van den Bosch, F.C., \& Balcells, M., 2003, ApJ, 583, 732
\reference Taffoni, G., Mayer, L., Colpi, M., \& Governato, F., 2003, MNRAS 341, 434
\reference Wadsley, J., Stadel J., \& Quinn T. 2003, astroph 0303521



\end{references}
\end{document}